# Bilateral and multilateral international scientific collaboration of EU member states: OpenAlex vs Scopus (2000-2024)

**Corresponding author:**
Myroslava Hladchenko, Centre for R&D Monitoring (ECOOM), University of Antwerp, Antwerp, Belgium
hladchenkom@gmail.com
Myroslava.Hladchenko@uantwerpen.be

**Abstract**
This study examines the evolution of bilateral and multilateral scientific collaboration among EU Member States and between the EU and global partners from 2000 to 2024 using data from OpenAlex and Scopus. The results show that OpenAlex, when restricted to cited articles, yields findings broadly comparable to those obtained from Scopus for assessing country-level research collaboration. Relative Intensity of Collaboration (RIC) values are consistently higher for multilateral than for bilateral partnerships. Increased collaboration intensity during the final years of FP7, the intermediate and later stages of Horizon 2020, and the final years of the study period suggests that EU FP may have strengthened collaboration among participating countries.

With regard to European integration, multilateral collaboration intensity increased between the EU-14 and EU-13, between these groups and EU candidate countries, and within the EU-13. Despite this growth, structural asymmetries persist. Bilateral collaboration among EU-14 countries is concentrated within the group and with EU-13, Brazil, Norway, Switzerland, and the United Kingdom, whereas EU-13 countries collaborate more intensively within the group, with EU candidate countries and Russia. EU-14 countries maintain stronger multilateral collaboration with high-income countries such as Australia, Canada, and the United States than do EU-13 countries. For both groups, collaboration with China remains the weakest. Although multilateral collaboration intensity with Russia has declined, it remained above the expected level for the EU-14 in 2024 and was 2.5 times higher than expected for the EU-13. This persistence may reflect the continued participation of Russian researchers in multilateral projects despite Russia's suspension from Horizon Europe in 2022.

## Introduction

The European Union's research and innovation strategy is grounded in the premise that Europe's long-term technological and economic competitiveness depends on a highly integrated innovation system, built on strong collaborative networks within the EU and with associated and candidate countries, as well as global research partners, alongside the broad diffusion of knowledge across borders (European Commission, 2020; 2024; COM, 2021; 2025). Scientific collaboration is widely recognised as a major driver of knowledge creation, research quality, and technological advancement (Adams, 2013; Wuchty et al., 2007). Examining collaboration within the EU and between the EU and global partners provides important insights into how national research systems are embedded in broader European and global scientific dynamics (Melin, 2000; Beaver, 2001; Persson et al., 2004). Understanding the intensity and structure of these cross-border linkages is therefore crucial for evidence-based policymaking, strengthening research capacity, and further developing the European Research Area (Makkonen & Mitze, 2016).

In Europe, scientific collaboration patterns have long been shaped by integration processes and research funding instruments at both EU and national levels (Glänzel & Schubert, 2001).
EU enlargements—together with Association Agreements for candidate countries—were designed not only to promote political and economic integration but also to align and strengthen research systems through participation in EU Framework Programmes (Luukkonen et al., 1992). FP7 (2007–2013), Horizon 2020 (2014–2020), and Horizon Europe (2021–2027) have played a central role in expanding multilateral collaboration both within the EU and globally, as their funding schemes emphasise multi-country consortia (Veugelers, 2021).

Contemporary global challenges increasingly require broad multilateral cooperation to mobilise diverse expertise and resources, and EU Framework Programmes explicitly encourage such collaboration to facilitate global responses to these challenges (European Commission, 2020; 2021; 2024). Despite the growth of multilateral collaboration, bilateral collaboration still dominates international scientific publishing (Adams & Gurney, 2018), partly because it is easier to organise and is strongly shaped by political, socioeconomic, geographical, cultural, and linguistic proximity (Hoekman et al., 2010; Scherngell & Barber, 2009).

The dual forces of European integration and globalisation raise important questions about how collaboration patterns evolve for countries at different stages of EU membership. Recent geopolitical disruptions, such as Brexit and Russia's invasion of Ukraine, could have implications for scientific collaboration of EU member states (Makkonen & Mitze, 2023).

At the same time, scholarly publishing has been undergoing a major transformation driven by Open Science and Open Access policies. This shift has contributed to the rise of OpenAlex as an openly accessible alternative to traditional proprietary bibliographic databases such as Scopus and Web of Science (Alperin et al., 2024; Culbert et al., 2025; Haunschild & Bornmann, 2024; Thelwall & Jiang, 2025). As a result, assessing how effectively OpenAlex captures country-level scientific collaboration, compared with Scopus, has become increasingly important for research evaluation and science-policy analysis.

This study has two main aims. First, it investigates how the Relative Intensity of Collaboration (RIC) for bilateral and multilateral research collaborations evolved for EU-14 and EU-13 member states within Europe and globally from 2000 to 2024, including the potential effects of major geopolitical disruptions such as Brexit and Russia's invasion of Ukraine. Collaboration partners include EU-14, EU-13, EC candidate countries, and major global research actors such as Australia, Brazil, Canada, Chile, China, Japan, India, Norway, South Africa, South Korea, Russia, Switzerland, the UK, and the USA. Second, the study examines whether OpenAlex provides results comparable to Scopus for assessing country-level collaboration patterns. The findings contribute to ongoing discussions on European research integration, science diplomacy, and the effectiveness of EU science policy (COM, 2021).

## EU Framework Pragrammes: EU integration and global collaboration

### EU integration

The rationale behind supporting research collaboration at the EU level is twofold: to enhance Europe's scientific excellence and strengthen European integration (Olechnicka et.al., 2019). The scientific integration of European states was encouraged through the launch of the European Framework Programme in 1984, by the European Communities (EC) and involved the ten European Economic Community member states. The EU enlargements of 2004 and 2007 contributed to the expansion of scientific collaboration with the Central and Eastern European countries that joined the EU, including Poland, the Czech Republic, Hungary, and the Baltic states. The share of internationally co-authored papers published by participants in the FP7 (2007–2013) increased by 11.5–11.9 percentage points in major funding streams (European Commission, 2015). Horizon 2020 introduced specific measures to strengthen the integration of Widening countries into the Framework Programme, targeting primarily the EU-13 and Associated Countries lagging behind (European Court of Auditors, 2022). In Horizon2020, the Widening countries also included Luxembourg and Portugal, while in Horizon Europe, they included Greece and Portugal from EU-14. Analysis of collaborative networks revealed that, compared with FP7, H2020 showed a tendency towards increased cooperation between EU-14 and EU-13 countries (EC, 2025). To further strengthen collaboration with Widening countries, the European Commission established the WIDERA (Widening Participation and Strengthening the European Research Area) component within Horizon Europe (EC, 2025). WIDERA calls typically require project consortia to be coordinated by institutions from Widening countries. Despite some progress in the success rate of Widening countries in obtaining funding through Horizon Europe, during the first two years of the programme, Germany alone proportionally received more than all the Widening countries. It implies that participation in EU programmes should be complemented by national governments reforming and strengthening the R&I sector (Kelk & Drake, 2023). There are dense clusters in co-authorship networks linking older and newer EU and associated members, often centred around Germany, France, Italy, Spain, and the UK as key collaboration hubs (European Commission, 2025**).** Germany has always been the central cooperation partner for Eastern European scientific communities (Glänzel & Schubert, 2004).

## EU Framework Programmes and encouragement of global collaboration

EU member states participate automatically in the Framework Programmes, whereas non-EU countries must conclude a Horizon Association Agreement to participate under equivalent conditions. Low- and middle-income countries may participate as partners and can receive EU funding, whereas high-income non-associated countries typically participate on a self-funded basis. Norway and Switzerland are associated with Horizon Europe. Despite Brexit and the transition period from 2016 to 2020, the UK remained associated with Horizon 2020 and was subsequently granted association to Horizon Europe, while continuing to participate in the European Cooperation in Science and Technology (COST) programme (Kelly, 2021). This continuity of association was expected to preserve established patterns of scientific collaboration between the UK and EU member states. By contrast, high-income countries such as Australia, Brazil, Canada, Chile, China, Japan, South Korea and the United States are not associated with Horizon Europe. There is a Brazil-EU MSCA Staff Exchanges scheme that is jointly funded by Brazil and Horizon Europe. The EU and Japan have also launched jointly funded calls under both Horizon 2020 and Horizon Europe (Annette, 2025). Chile has participated in EU research programmes as an international partner, engaging in projects under FP7 and Horizon 2020 and receiving approximately €9.7 million in funding under the latter programme, and continues to collaborate under Horizon Europe, typically on a self-funding basis unless its participation is deemed essential (EEAS 2023). Jointly funded calls with China were also implemented in FP7, Horizon 2020 and Horizon Europe (European Commission, Directorate-General for Research and Innovation, n.d.). However, at the end of 2025, the European Commission announced the exclusion of Chinese universities, including the so-called "Seven Sons of National Defence," from participating in roughly half of Horizon Europe, particularly in clusters related to health, digital technologies, and civil security, citing security and dual-use concerns (Matthews, 2025). India and South Africa participate as international partners and are eligible for EU funding. Despite the absence of full association, collaboration with these countries continues to deepen: in 2025, the EU and India launched joint calls worth approximately €60 million (Euraxess, 2024). In 2022, the European Commission suspended Russian public bodies from participation in Horizon 2020 and Horizon Europe projects, terminating their involvement in ongoing grants and halting payments to these entities (EC, 2022)

## Bilateral vs multilateral collaboration

EC prioritises multilateralism in science collaboration to facilitate global responses to global challenges (COM, 2021). Multilateral collaboration offers numerous benefits, allowing national governments to overcome financial constraints by granting access to advanced instrumentation required for cutting-edge research (Lima-Toivanen et al., 2025; Burke, 2025). International science is becoming increasingly networked, and large global teams are expanding at the quickest rate (Adams & Gurney 2018; Wagner et al. 2017; Wuchty et al. 2007).

Apart from multilateral research projects, EC also admits the importance of bilateral collaboration (COM, 2021). Bilateral scientific collaboration is typically formalised through memoranda of understanding (MoUs) or Science, Technology, and Innovation (STI) cooperation agreements between the European Union (EU) and non-EU countries, or between individual EU member states and external partners, for example between the EU and the USA (European Commission, 2023), EU and South Korea (EC & Republic of Korea, 2007).

Despite the rapid global rise of multilateral collaborations, bilateral partnerships continue to dominate international scientific publishing, largely because two-country collaboration is easier to organise and it remains the most common form of cross-border co-authorship (Glänzel & Schubert, 2004; Wagner & Leydesdorff, 2005; Gazni, Sugimoto & Didegah, 2012; Adams & Gurney, 2018).

The selection of a collaboration partner for both bilateral and multilateral collaboration depends on several factors. First, countries tend to collaborate more intensively with partners that are similar in terms of scientific capacity and economic development. This applies both to neighbouring countries and to global collaborations, reflecting a broader pattern in which socioeconomic, political, geographical, cognitive, linguistic, and cultural proximities foster scientific cooperation (Hoekman et al., 2009; Glänzel & Schubert, 2004; Zitt et al., 2000; European Commission, 2025; Kwiek, 2021). As a result, core scientific countries tend to collaborate more intensively with one another than with countries outside the scientific core (Hoekman et al., 2009). Second, R&D funding policies and research capacity, including research infrastructure, also influence with whom and how countries collaborate (European Commission, 2025; Kwiek, 2021).

### *Trends in global collaboration*

From a global perspective, the USA occupies a central position in the worldwide scientific network (Olechnicka et al., 2019; Gui et al., 2019; Marini & Mouritzen, 2025), serving as the largest international collaborative partner for countries including China, the UK, Germany, France, Italy, Spain, the Netherlands (Kwiek, 2019) and South Korea (Gueye et al., 2022). Overall, the USA, U.K., Germany, France, Italy, and Canada are the core countries in the global collaboration network, collectively accounting for 82% of multinational publications (Gazni et al., 2012). In Canada, the USA is the primary partner of all provinces, largely due to geographical proximity; however, researchers from the French-speaking province of Quebec maintain stronger collaborative ties with scholars in France and Belgium (Larivière et al., 2004). South Korean researchers have traditionally collaborated more extensively with colleagues in the USA than with those in the EU (Gueye et al., 2022).

Since the first decade of the 21st century, alongside the USA, China has gained a prominent position as the central hub of the global scientific collaboration network **(**Veugelers, 2017**).** Germany, the UK, and the Netherlands are among China's top European collaborators (Gómez-Espés et al, 2024; Wang et al., 2017). Scientific collaboration between the European Union (EU) and Latin America has expanded steadily over the past two decades, driven by shared priorities in research and innovation (Belli & Nenoff, 2022) and the EU science diplomacy policy (Uribe-Mallarino, 2022). Spain and Portugal have historically been the strongest partners of Latin America due to linguistic ties and historical connections (Chinchilla-Rodríguez et al., 2016; Russell et al., 2020; Lemarchand, 2012; Uribe-Mallarino, 2022; Gueye et al., 2022). Overall, scientific ties between the EU and Latin America reflect both regional priorities and the growing need to address global challenges through collaborative research (Ronda-Pupo, 2024).

For scholars from sub-Saharan Africa, co-authorship with European scholars is the second most common form of collaboration (5.7%), following intra-regional collaborations. This can be attributed to colonial ties (Gueye et al., 2022). This is also the case for South African scholars, who prioritise collaboration with colleagues from the USA, the UK, Germany, Canada, Australia, France, the Netherlands, Italy, and Belgium (Sooryamoorthy 2009; Heleta & Jithoo, 2023; Pouris & Ho, 2014).

## Data and Methodology

### Data

The countries in question in this study are EU member states, which include EU-14 (Austria, Belgium, Denmark, Finland, France, Germany, Greece, Ireland, Italy, Luxembourg, Netherlands, Portugal, Spain, Sweden) and EU-13 (Bulgaria, Croatia, Cyprus, Czech Republic, Estonia, Hungary, Latvia, Lithuania, Malta, Poland, Romania, Slovakia, Slovenia). The collaborative partners include EU-related countries like EU-14, EU-13, EU candidates (Albania, Bosnia and Herzegovina, Georgia, Moldova, North Macedonia, Serbia, Turkey, Ukraine) and global partners:

Australia, Brazil, Canada, Chile, China, India, Japan, Norway, Russia, South Africa, South Korea, Switzerland, the UK, and the USA.

Data for this study were taken from the CWTS Scopus in-house database and OpenAlex, covering articles published between 2000 and 2024. For OpenAlex, only articles with at least one citation were included to ensure that the dataset reflects publications that have had measurable impact and are actively engaged in the scholarly communication network, thereby reducing noise from uncited or less influential works. This approach also mitigates potential distortions associated with incomplete affiliation metadata in OpenAlex. Hladchenko (2026) revealed that around 30% of authors per year have incomplete affiliation metadata in OpenAlex, with this percentage rising substantially in recent years. The approach of including only cited articles from OpenAlex follows Engels et al. (2026).

The Relative Intensity of Collaboration (RIC) (Fuchs et al., 2021) for bilateral and multilateral collaboration pairs was calculated using the CWTS Scopus in-house database and on BigQuery using OpenAlex data stored by the INSYSPO project.

A key limitation of this study is that it relies on the CWTS in-house version of Scopus rather than the public Scopus interface or API. The RIC indicators can not be calculated directly from Scopus for the full 2000–2024 period due to technical and data coverage constraints. The Scopus web interface and API impose maximum retrieval limits (typically up to 2,000 documents via the interface and 5,000 via the API), which are easily exceeded when working with large-scale longitudinal datasets. Furthermore, export formats often do not include complete lists of authors and affiliations for publications with many co-authors, which are essential for accurately measuring multilateral collaboration. These practical restrictions make it infeasible to construct a comprehensive, structured dataset covering the entire period directly from Scopus.

While the CWTS database is based on Scopus data, it is preprocessed, standardised, and internally structured, which may result in some differences in record coverage, metadata structure, or update frequency compared with direct Scopus access. Consequently, exact replication of the results by external researchers using the standard Scopus interface may not be possible. Nonetheless, the CWTS database is widely used in scientometric research, and its coverage and data quality are comparable to the original Scopus source, ensuring the validity of the analysis.

Overall, this limitation highlights the advantages of OpenAlex for large-scale historical collaboration analyses.

**Relative Intensity of Collaboration (RIC)**

Relative Intensity of Collaboration (RIC)**,** a widely used metric for quantifying the strength of research collaborations (Coccia & Bozeman, 2016; Fuchs et al., 2021; Luukkonen et al., 1992). This study uses RIC as the primary measure to assess the intensity of bilateral and multilateral international collaborations, allowing a systematic comparison of EU member states' collaborative activity across different partner countries and country groups.

The Relative Intensity of Collaboration (RIC) (Fuchs et al., 2021) was used as a main metric in this study. The formula for RIC was taken from Fuchs et al. (2021). In the RIC formula:

- Cxy — the number of collaborative publications between entities X and Y;
- Cx — the total number of publications involving entity X;
- Cy — the total number of publications involving entity Y;
- T — the total number of publications in the analyzed network.

The RIC indicator compares the share of country Y within the collaboration profile of country X to the share of Y within all collaborations in the whole network. The Relative Intensity of Collaboration (RIC) is calculated as the ratio of the observed share of co-authored publications between two countries to the expected share based on their total publication outputs (Fig. 1). The RIC indicator is asymmetric and increases in value if collaboration between the two countries increases. In this study, the publication-based interpretation of the RIC indicator developed by Fuchs et al. (2021) was used. RIC values of 1.0 indicate that the collaboration between two countries occurs as frequently as expected, based on their total publication outputs and the overall network structure, essentially reflecting random mixing. A value greater than 1.0 suggests preferential or stronger-than-expected collaboration, while a value below 1 indicates weaker-than-expected ties. This approach aligns with established practices in bibliometric studies of international co-authorship (Katz & Martin, 1997).

|  | not Y | Y |
|---|---|---|
| X | $C_X - C_{XY}$ | $C_{XY}$ |
| not X | $T - C_X - (C_Y - C_{XY})$ | $C_Y - C_{XY}$ |

$$RIC(X,Y) := \left(\frac{C_{XY}}{C_X}\right) / \left(\frac{C_Y - C_{XY}}{T - C_X}\right) = \frac{C_{XY} \cdot (T - C_X)}{C_X \cdot (C_Y - C_{XY})}$$

Fig. 1. Diagram and formular of RIC (Fuchs et a., 2021).

**Statistical tests**

To examine the evolution of RIC for bilateral and multilateral research collaborations and the difference between Scopus and OpenAlex from 2000 to 2024, an interaction-based multiple linear regression models were fitted using the formula RIC ~ publication_year+ group * collaboration_patner + collaboration_type + database, $\Delta$RIC ~ publication_year+ group * collaboration_patner + collaboration_type. The models were estimated in R using the stats package (lm function), with data manipulation performed using the dplyr package. Interaction terms between group and collaboration partner allowed assessment of group-specific temporal trends in RIC.

Additionally, separate regression models were fitted to explore the evolution of RIC for bilateral and multilateral collaboration for the EU-14 and EU-13 with Russia, China and the USA (2013-2024), and the UK (2016-2024).

Peaks in RIC from 2010 to 2024 were identified by selecting the top two RIC values per group (UE-14 and EU-13), collaboration partner, publication type, collaboration type and database, retaining ties to account for multiple high-impact publications. The analysis was conducted in R using the dplyr package for data manipulation and ggplot2 for visualisation.

## Results

### Coverage of bilateral and multilateral collaborations in OpenAlex and Scopus

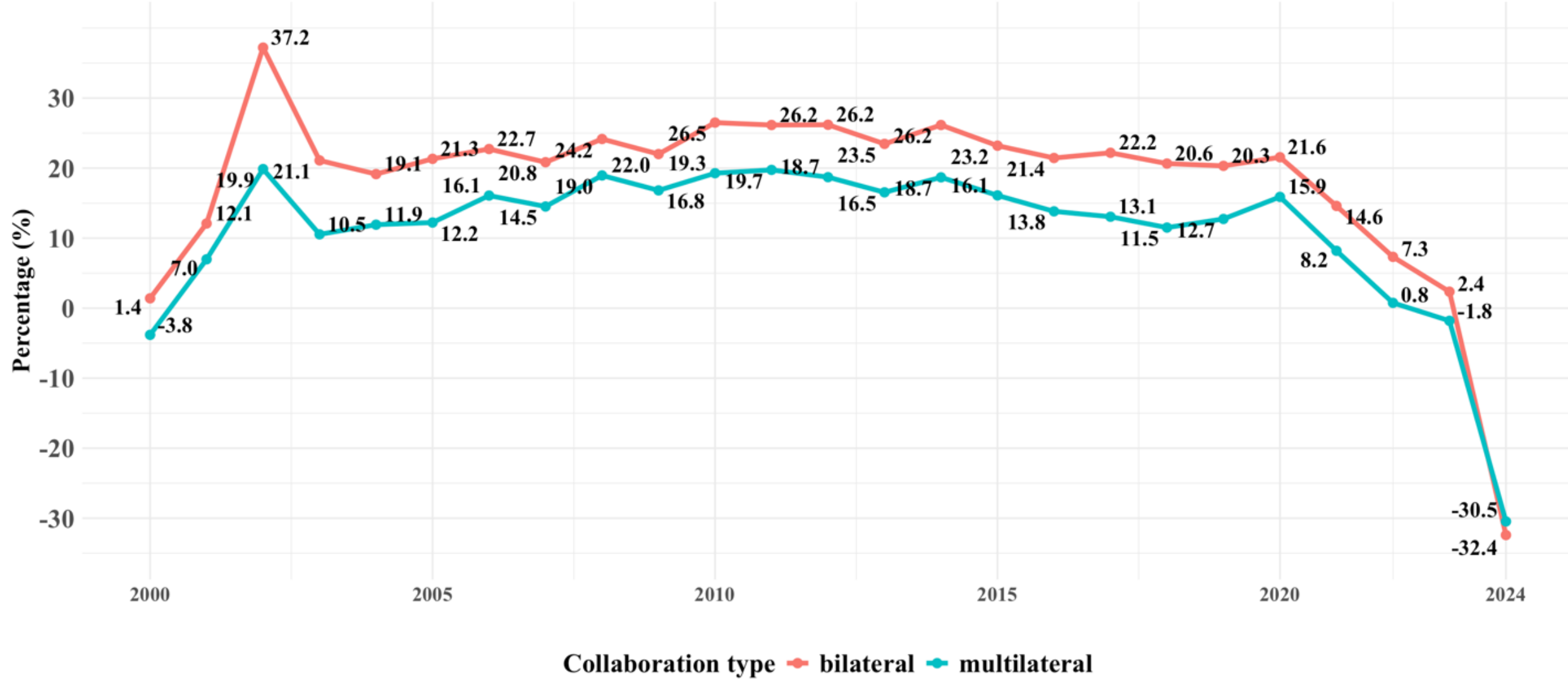

**Fig. 1 Percentage excess of OpenAlex coverage compared to Scopus**

Figure 1 illustrates the percentage by which OpenAlex coverage exceeds Scopus for bilateral and multilateral collaborations from 2000 to 2024. Both collaboration types show generally higher OpenAlex coverage, particularly for bilateral articles, with the difference peaking at 37.2% in 2002. After 2020, both lines drop sharply, reaching negative values in 2024 (−30.5% and −32.4%). This decline is attributable to the inclusion of only OpenAlex articles that had received at least one citation, which disproportionately excludes recently published articles that have not yet accumulated citations. OpenAlex covers more bilateral than multilateral international collaborations compared to Scopus, because many bilateral articles appear in journals not indexed by Scopus.

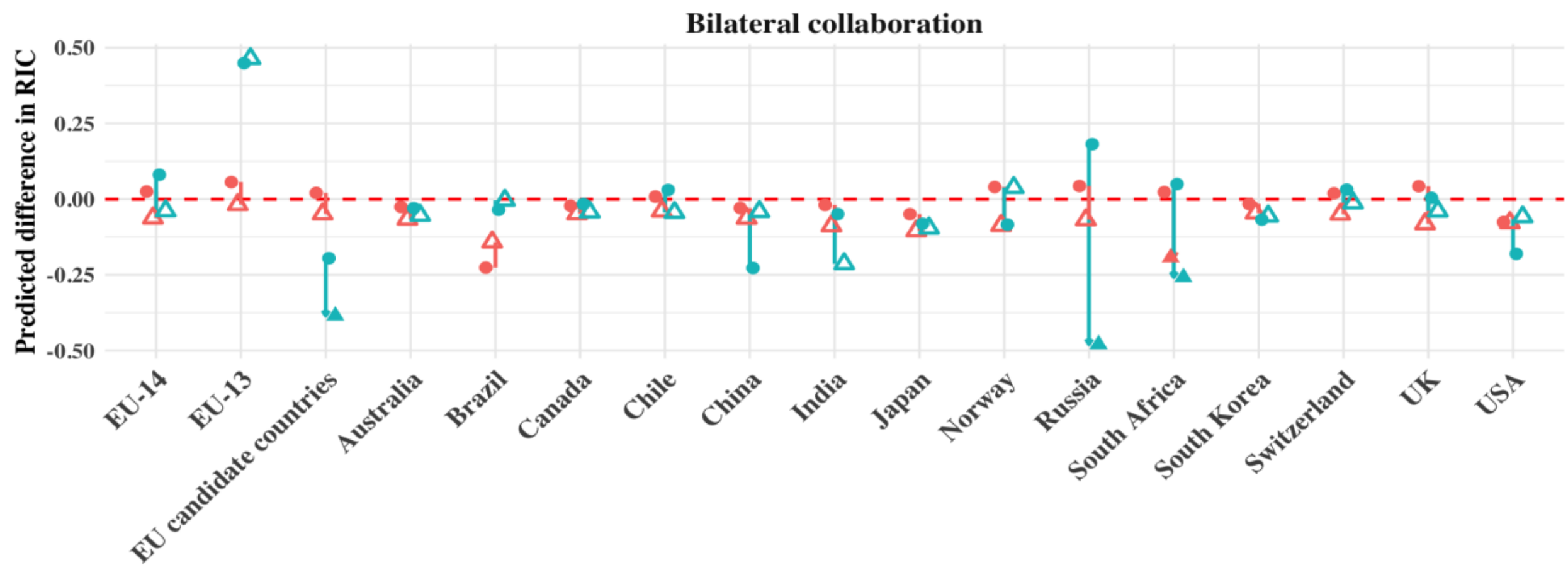

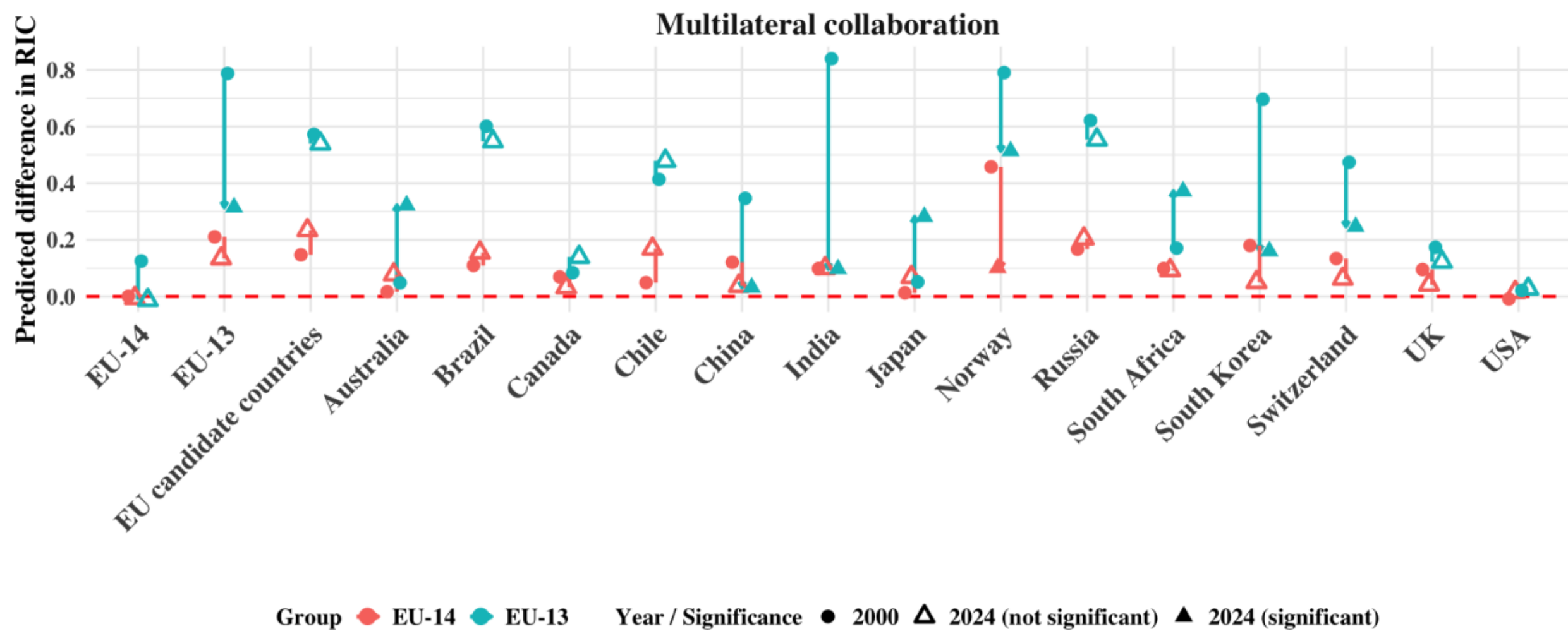

**Fig. 2** Predicted ΔRIC (Scopus–OpenAlex by collaboration partner 2000 vs 2024

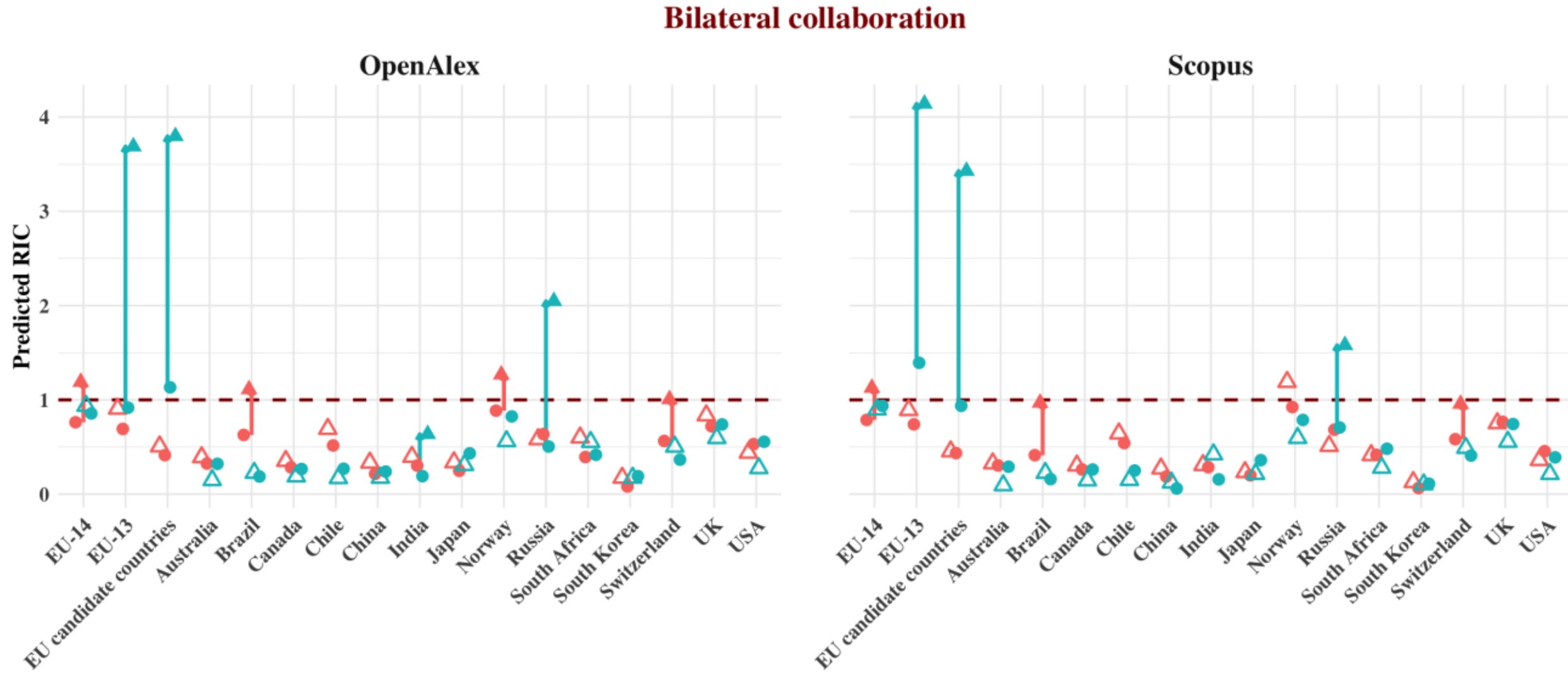

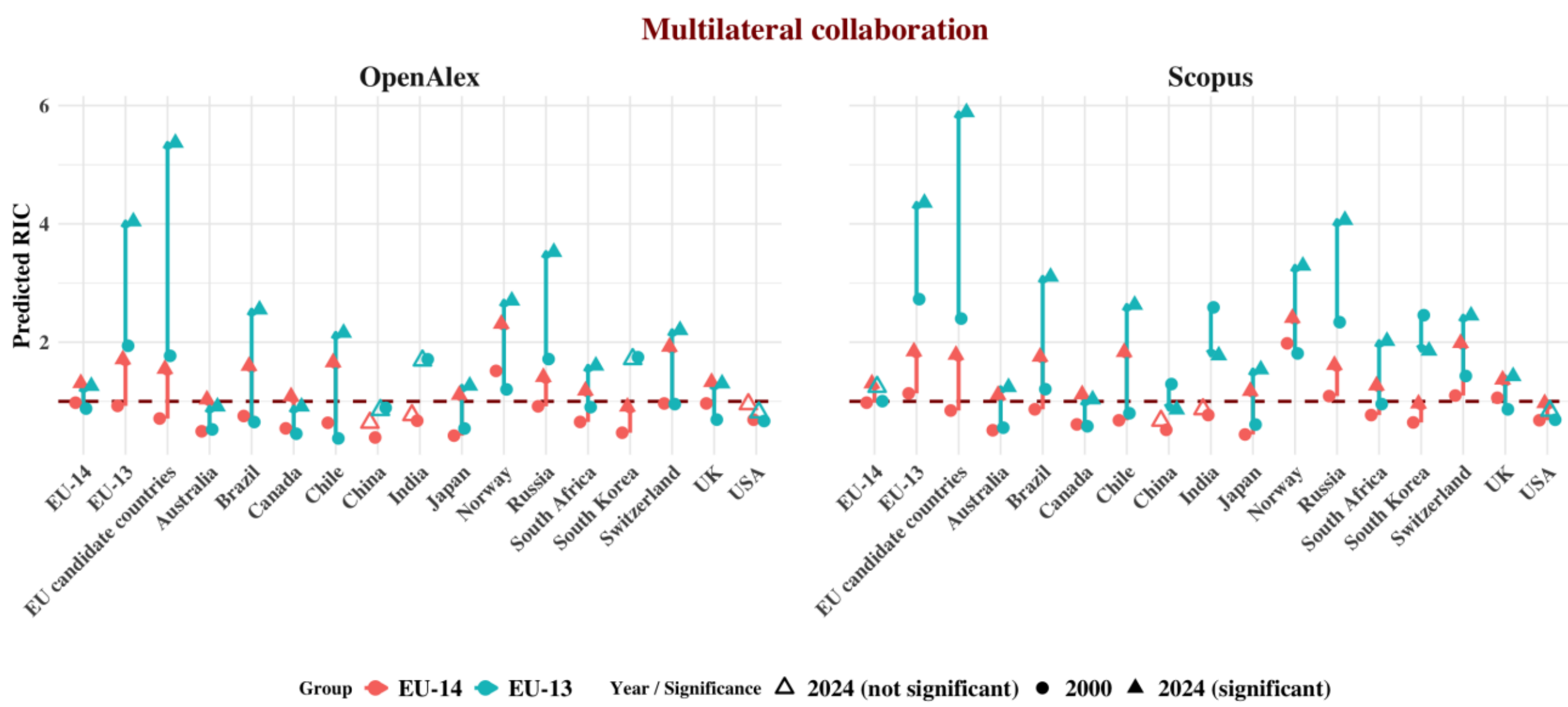

**Fig. 3** Predicted RIC by collaboration partner 2000 vs 2024 in OpenAlex and Scopus

The predicted ΔRIC (Scopus – OpenAlex), derived from a multiple regression model incorporating a fully interacted four-way structure among publication year, EU group, collaboration partner, and collaboration type, is presented in Fig. 2. Although the model is statistically significant (adjusted $R^2 = 0.207$, $F(135, 21584) = 43.03$, $p < 2.2\times10^{-16}$), it explains only about 21% of the variation, indicating that the included variables have limited explanatory power for ΔRIC.

For bilateral collaboration, ΔRIC with most partners remains close to zero and relatively stable over time. In several cases, OpenAlex reports slightly higher RIC than Scopus. For multilateral collaboration, Scopus tends to report marginally higher RIC than OpenAlex, although the differences are small and generally decreasing. For multilateral collaboration, EU-13 countries show consistently larger positive ΔRIC with partners such as Brazil, Norway, China, South Korea, India, and Switzerland, with coefficients ranging from +0.14 to +0.70 (all $p < 0.05$). Across both collaboration types, larger deviations are observed for EU-13 than EU-14 countries.

Figure 3, using OpenAlex data, shows that within EU-related countries, for bilateral collaboration in 2024, EU-14 countries have the highest predicted RIC within the group and the lowest with EU candidate countries. Collaboration with both EU-13 and EU candidate countries remains below the expected level. In contrast, for multilateral collaboration, EU-14

countries exhibit the highest predicted RIC with EU-13 and the lowest within the group. In all three cases, the predicted RIC in 2024 exceeds the expected level. For both bilateral and multilateral collaboration, EU-13 countries have the highest predicted RIC in 2024 with EU candidate countries and the lowest with EU-14. In all three cases, predicted RIC in 2024 is above the expected level.

From a global perspective, for bilateral collaboration, EU-14 countries have the highest predicted RIC in 2024 within the group and with Brazil, Chile, Norway and Switzerland, whereas EU-13 countries have the highest predicted RIC also within the group, as well as with EU candidate countries and Russia. For multilateral collaboration, the predicted RIC values in 2024 are higher than for bilateral collaboration and exceed the expected level with most partners. For EU-14 countries, the predicted RIC in 2024 does not reach 1.0 with China, India, South Korea and the USA (RIC=0.95, not statistically significant but significant according to Scopus). For EU-13 countries, the predicted RIC in 2024 remains below 1.0 with China, Australia (0.91), Canada (0.91), and the USA (0.82). The predicted RIC with South Africa in 2024 is above the expected level for both types of collaboration among the EU-14 and EU-13.

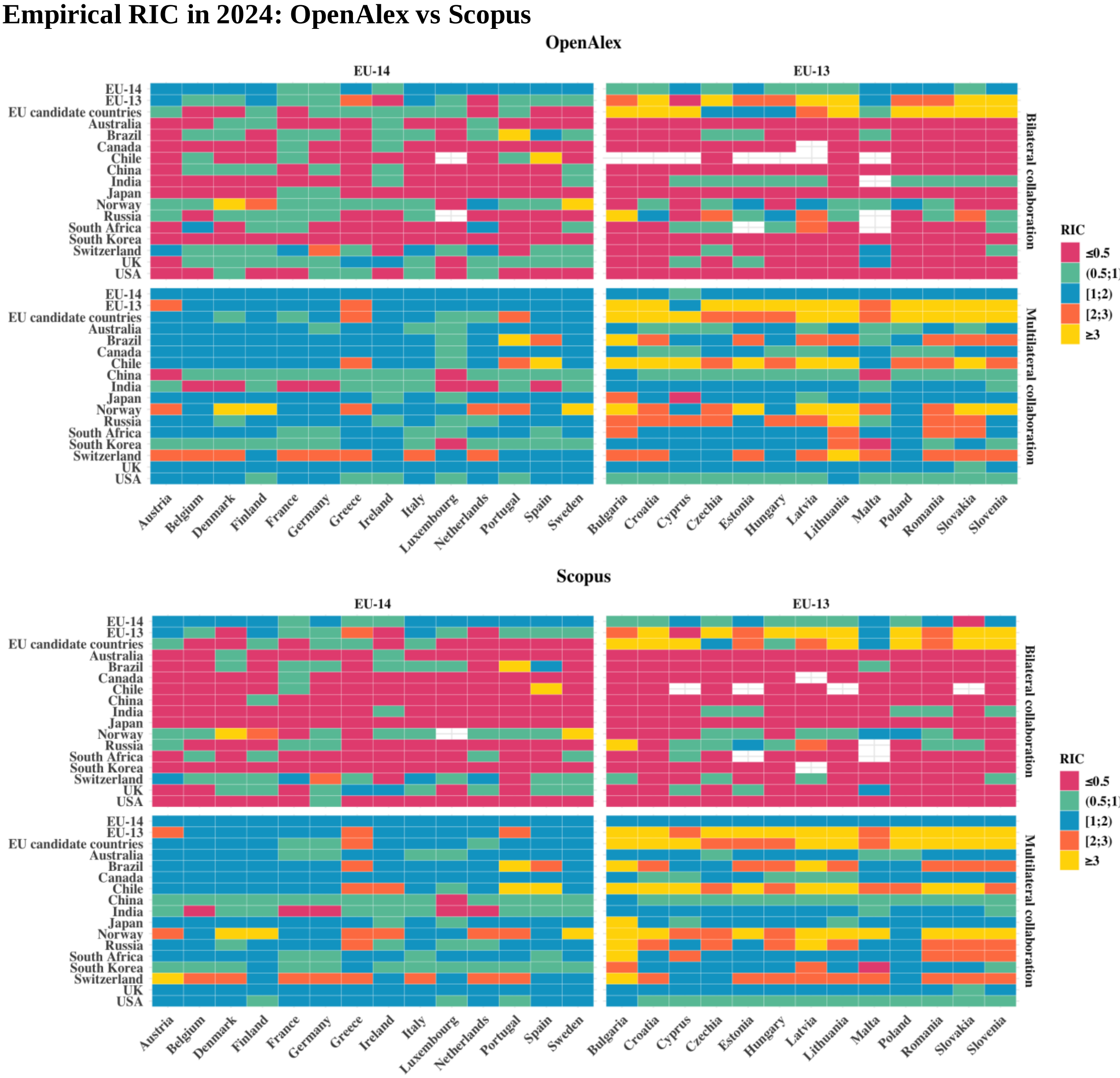

**Fig. 4** Empirical RIC in 2024: Comparison between OpenAlex and Scopus

Figure 4 compares empirical RIC in 2024 between OpenAlex and Scopus. Overall, this figure shows that EU-14 countries have lower bilateral RIC values in Scopus than in OpenAlex, as indicated by a broader presence of red cells in the Scopus heatmap. In contrast, EU-13 countries display higher RIC values in Scopus than in OpenAlex.

The OpenAlex heatmap highlights clear differences in collaboration patterns between EU-14 and EU-13 countries. For bilateral collaboration, EU-13 countries predominantly exhibit RIC values $\leq 0.5$, whereas EU-14 countries more frequently show values above 0.5 and up to 1.0. EU-14 countries generally show higher RIC values with high-income countries like Australia, Norway, Switzerland, the UK, and the USA, whereas EU-13 countries display stronger ties with EU candidate countries and Russia.

Multilateral collaboration produces higher and more evenly distributed RIC values for both groups, with EU-13 countries exhibiting particularly stronger engagement than EU-14 with EU candidate countries, Chile, and Russia. EU-14 collaboration remains concentrated among high-income countries, including Brazil and Chile. For the EU-14, RIC values within the group and with EU-13 countries exceed the expected level. The lower intensity is observed with EU candidate countries. EU-14 countries have the lowest RIC values for multilateral collaboration with China, India, and South Korea, whereas higher RIC values are observed with South Africa. EU-13 countries exhibit a higher intensity of multilateral collaboration with India and South Korea than EU-14 countries, but a lower intensity of collaboration with Australia, Canada, and the USA.

For both bilateral and multilateral collaboration, countries exhibit higher RIC with neighbouring countries, for example, RIC above the expected level have Austria, France, Germany and Italy with Switzerland and Denmark, Finland, and Sweden with Norway. The former colonial ties of Belgium and the Netherlands with South Africa are reflected in bilateral collaboration, RIC values that exceed the expected level. Among EU-14 countries, Greece has the highest RIC with EU-13 and EU candidate countries and Portugal with EU candidate countries, which can be attributed to Greece and Portugal being categorised as Widening countries and being involved in WIDERA.

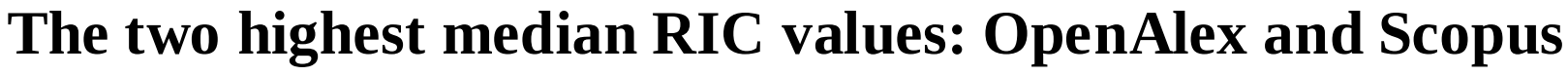

**The two highest median RIC values: OpenAlex and Scopus**

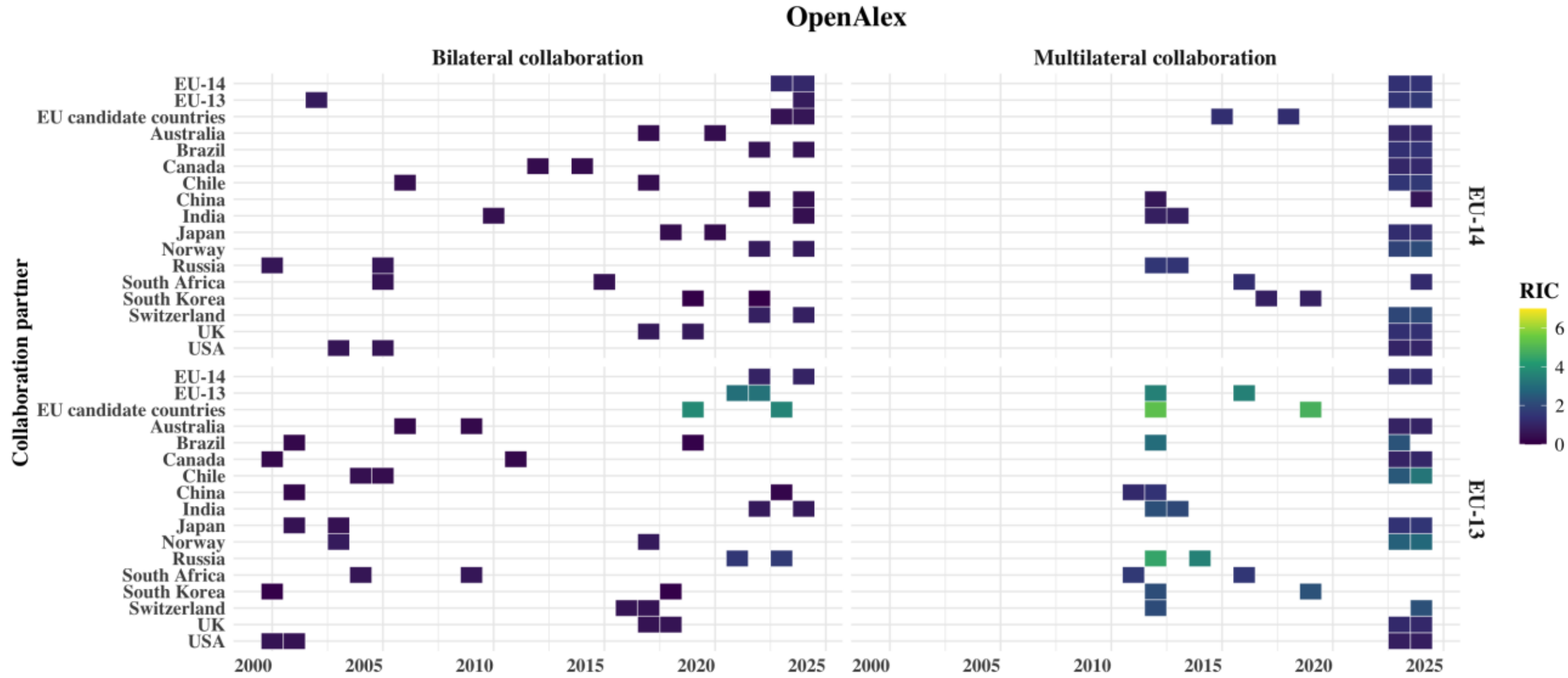

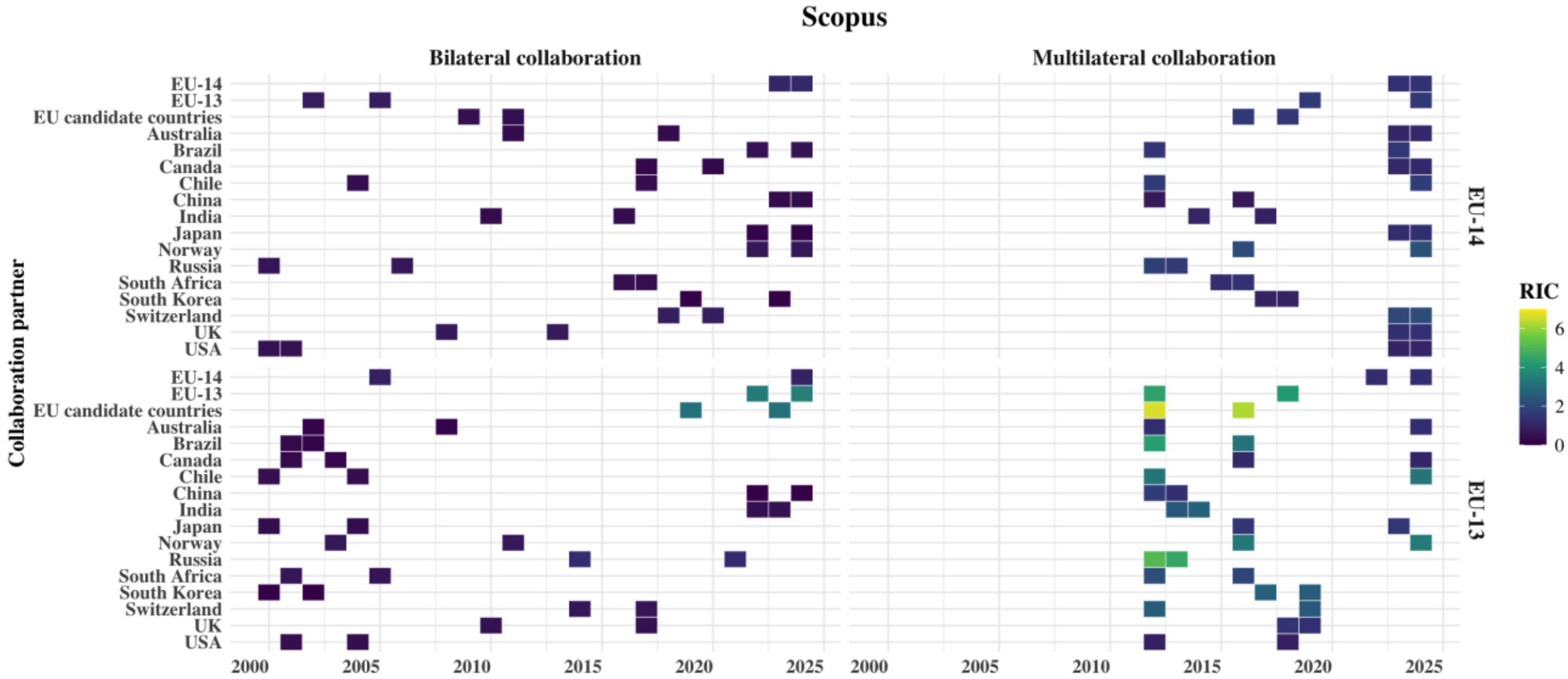

**Fig. 5 The two highest median RIC values by collaboration partner: OpenAlex and Scopus**

Figure 5 shows the two highest median RIC values by collaboration partner during 2010–2024 according to OpenAlex and Scopus. Based on OpenAlex data, bilateral collaboration among EU-14 countries exhibits a concentration of peak years in 2022–2024. These peaks are observed for collaboration within EU-14, as well as with EU-13 countries, EU candidate countries, Brazil, China, Norway, and Switzerland. In contrast, peak years for EU-13 countries are distributed between the beginning and the end of the study period. The most recent bilateral peaks (2022–2024) are observed for collaboration with EU-14 countries, EU-13 countries, EU candidate countries, India, and Russia.

For multilateral collaboration, both groups display a broader temporal distribution of peak years. Several peak years occur around 2012–2013 and 2017–2019. The former period coincides with the conclusion of FP7, whereas the latter falls within the implementation period of Horizon 2020. A notable concentration of peak years is also observed in 2023–2024.

The Scopus heatmap for bilateral collaboration broadly mirrors the patterns observed in OpenAlex. For multilateral collaboration, however, peak years are more frequently concentrated between 2016 and 2019, which corresponds to the intermediate and later years of Horizon 2020, suggesting some differences in the timing of maximum collaboration intensity between the two databases.

### Russia’s invasion of Ukraine and collaboration of EU-14 and EU-13 with Russia, China and the USA (2013-2024)

The effects of Russia’s invasion of Ukraine on collaboration patterns of EU member states were explored in the context of their engagement with such major global scientific systems as the USA and China. Figure 6 compares predicted RIC in OpenAlex and Scopus for EU-14 and EU-13 with Russia, China, and the USA from 2013 to 2024. According to OpenAlex, bilateral collaboration between EU-14 and Russia declines and remains below the expected level, whereas EU-13 shows a marked increase, with RIC reaching twice the expected value; both trends persist after 2022. Scopus indicates a slight decrease for both EU-14 and EU-13. For multilateral collaboration, both databases show a decreasing RIC with Russia for EU-14 and EU-13 countries; however, in 2024, collaboration intensity remains above the expected level for EU-14 and is approximately 2.5 times higher than expected for EU-13.

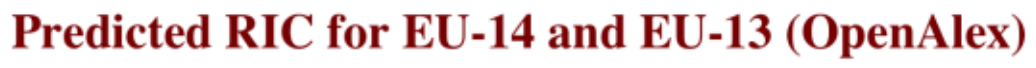

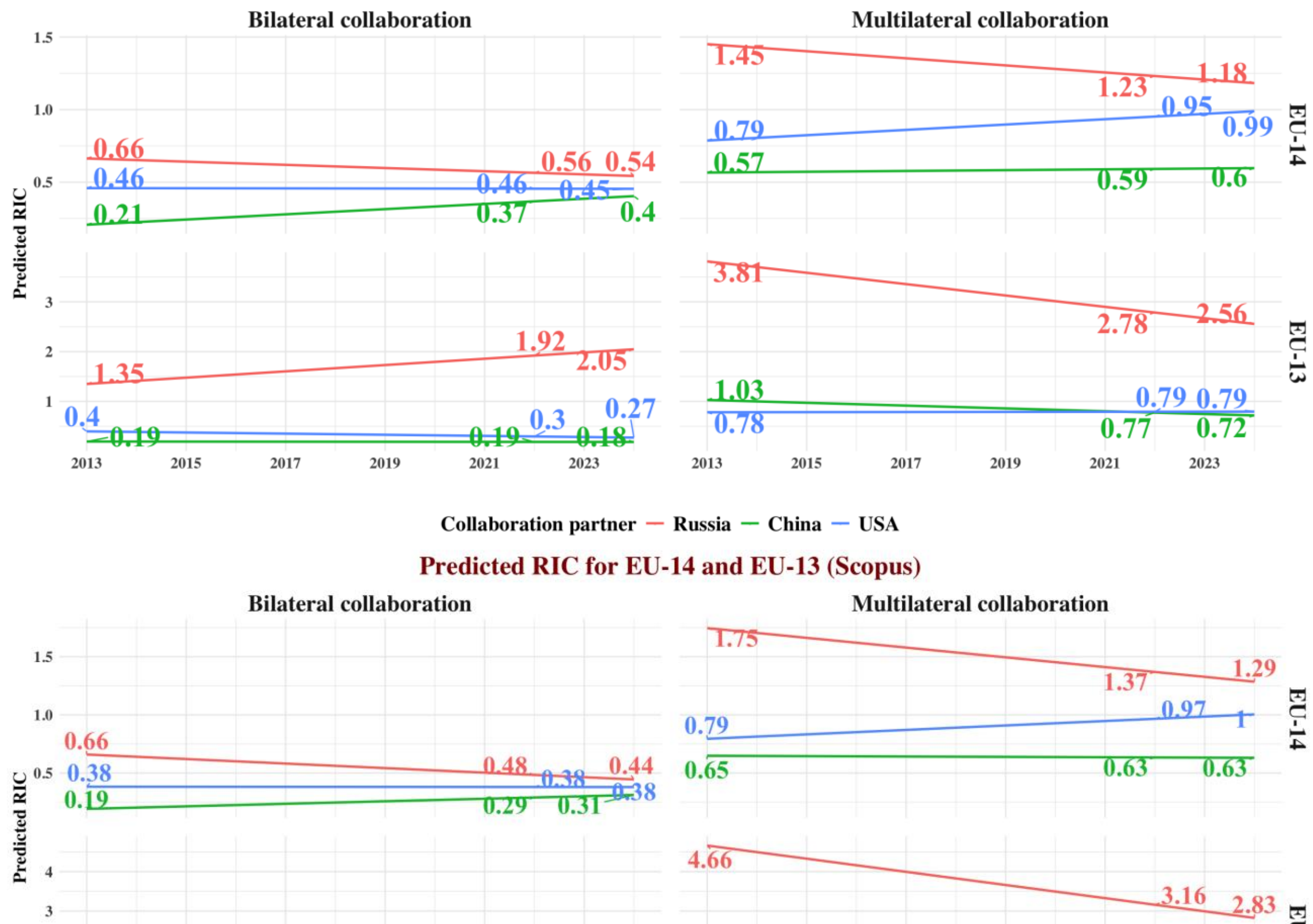

**Fig. 6** Predicted RIC for EU-14 and EU-13 with Russia, China and the USA (2013-2024)

With China, OpenAlex shows stable bilateral collaboration for EU-13 and a slight gradual increase for EU-14, although in both cases, the RIC values do not reach even half of the expected level. Multilateral collaboration for EU-14 remains virtually unchanged in both databases and far below the expected level. For EU-13, multilateral RIC with China declines steadily and falls below the expected level.

With the USA, bilateral RIC for EU-14 remains stable across databases, whereas EU-13 experiences a decrease to low RIC levels by 2024. Multilateral RIC increases for EU-14 and almost reaches the expected level, but decreases for EU-13, being below the expected level.

Figure 7 illustrates predicted RIC for EU-14 and EU-13 with the UK in OpenAlex and Scopus from 2016 to 2024. It reveals a decline in bilateral collaboration with the UK, more pronounced for EU-13 than EU-14. For multilateral collaboration, there is an upward trend for EU-14, with RIC being above the expected level already in 2016. In contrast, for the EU-13, the intesntity of collaboration decreases, though RIC is above the expected level.

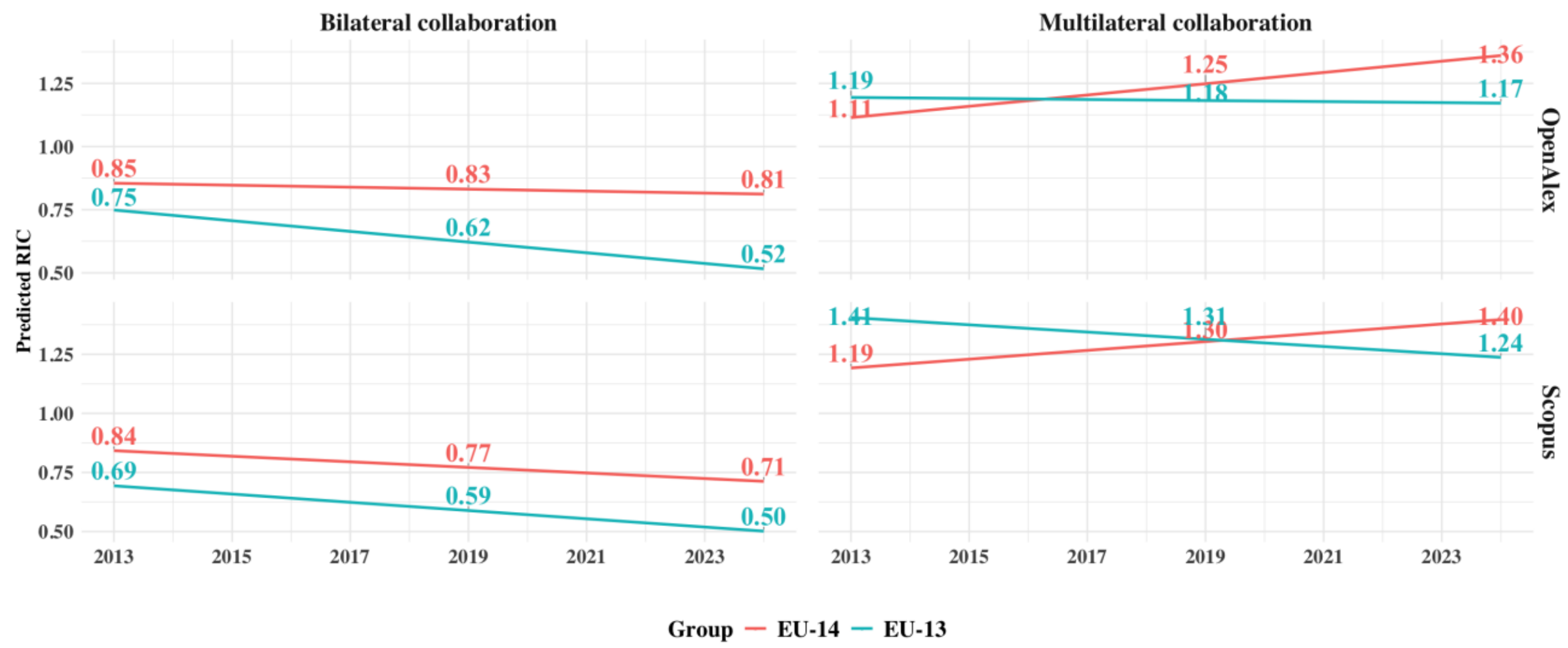

**Fig. 7** Predicted RIC for EU-14 and EU-13 with the UK (OpenAlex vs Scopus, 2016-2024)

## Conclusions

### *OpenAlex vs Scopus*

To summarise, the findings demonstrate that OpenAlex, when restricted to cited articles, yields results that are broadly comparable to those obtained from Scopus for the assessment of country-level research collaboration. Differences in RIC values between the two databases are more pronounced for EU-13 than for EU-14. A key advantage of OpenAlex is that RIC can be calculated directly using BigQuery, a feature not available in Scopus. That is why this study relies on the CWTS in-house version of the Scopus database rather than the Scopus platform itself. This may constitute a limitation, as differences in data processing, updating procedures, and record coverage could lead to discrepancies between the two data sources.

### *Relative intensity of collaboration from 2000 to 2024*

Relative Intensity of Collaboration (RIC) has grown strongly over the past 25 years, both bilaterally and multilaterally. Multilateral collaboration is consistently stronger than bilateral collaboration, highlighting the importance of international research consortia targeting global challenges (Adams & Szomszor, 2022; Hsiehchen et al., 2015; Lima-Toivanen et al., 2025; Burke, 2025). Based on empirical data for 2024, the relative intensity of multilateral collaboration with most partners is either slightly below or above the expected level.

### *Intensity of collaboration as a result of mutual initiatives of the EU and partners*

Collaboration intensity results from initiatives undertaken by the EU, its member states, and their collaboration partners. From the EU side, the results indicate an increase in collaboration intensity during the final years of FP7, the intermediate and later stages of Horizon 2020, and the final years of the study period, a pattern that is consistent with observations reported by the European Commission (2015). Overall, the findings reveal that EU Framework Programmes may have contributed to strengthening collaboration intensity among EU-related countries as well as with global partners. This interpretation is consistent with previous studies that highlighted the role of Framework Programmes in fostering scientific collaboration and network integration (Hoekman et al., 2013; Balland et al., 2019). The United Kingdom remains an important research partner for both groups, with RIC values exceeding the expected level. This pattern likely reflects the UK's continued association with Horizon Europe despite Brexit.

From a global perspective, an increase in the intensity of multilateral collaboration with Brazil, Chile, and Japan demonstrates the results of mutual research funding initiatives in general and within Horizon in particular. The strong ties observed between EU member states and Brazil and Chile are consistent with earlier findings (Belli & Nenoff, 2022), which show that EU–Latin America scientific collaboration has expanded steadily over the past two decades, driven by shared research and innovation priorities. Although multilateral collaboration between the EU-14 and South Korea remains slightly below the expected level, South Korea's association with Horizon Europe from 2025 onward is expected to further strengthen research collaboration between the two parties.

Drawing on empirical RIC data for 2024, EU-14 countries have the lowest intensity of multilateral collaboration with China, India, and South Korea, whereas EU-13 countries show the lowest intensity with Australia, Canada, and the USA.

As regards China, EU-14 has shown a consistently low and stagnant intensity of multilateral collaboration, while EU-13 has exhibited a declining trajectory relative to expected collaboration levels. However, in 2013, the predicted RIC between EU-13 countries and China was at the expected level, consistent with Wang et al. (2017), who reported a substantial strengthening of collaboration between EU-13 countries and China from 2000 to 2014.

***Asymmetries in collaboration patterns or regional/economic clustering***

Asymmetries are observed in collaboration patterns of both EU-14 and EU-13. For EU-14, bilateral collaboration is concentrated mainly within the group and with EU-13 and economically similar neighbours: Norway, Switzerland. Regarding multilateral collaboration, the EU-14 also shows strong collaboration with other high-income countries such as Australia, Canada, Brazil, Chile, and the United States. Most EU-14 countries exhibit multilateral collaboration intensity with the United States above the expected level, reaffirming their central position in transatlantic scientific cooperation, consistent with Kwiek (2021). EU-13 collaborate above the expected level within the group, with EU candidate countries, and Russia, which aligns with Hladchenko (2026a). Multilateral collaboration between the EU-14 and the USA has strengthened, but it remains unchanged between the EU-13 and the USA. These collaboration patterns of the EU-14 and EU-13 align with prior findings showing that countries tend to collaborate with others at similar levels of scientific capacity or economic development (Hoekman et al., 2009).

Collaboration patterns of the EU-14 and EU-13 also reflect the influence of geographical, cultural and linguistic proximity. This is evident in the following partnerships, which exhibit high collaboration intensity: Ireland with the UK; Germany, France, and Italy with Switzerland; Denmark, Finland, and Sweden with Norway; France with Canada; Spain with Chile; Portugal with Brazil. These patterns align with global collaboration trends reported by prior studies (Larivière et al., 2004; Chinchilla-Rodríguez et al., 2016; Russell et al., 2020; Lemarchand, 2012; Hoekman et al., 2009). The Netherlands and Belgium show strong collaboration with South Africa, supporting prior studies on the impact of historical colonial ties on scientific collaboration (Gueye et al., 2022).

***RIC trajectory patterns***

RIC trajectories reveal that EU-13 countries exhibit greater variability, with more frequent extreme deviations in both positive and negative directions, compared with the more stable patterns observed for the EU-14. Such volatility suggests a less consolidated or more externally sensitive research environment. In contrast, RIC values clustering around 1.0—more common among EU-14 countries—reflect characteristics of a well-funded and internationally embedded scientific system, where collaborative activity is balanced and structurally integrated within global research networks.

***Collaboration with Russia following the full-scale invasion of Ukraine***

Although multilateral collaboration with Russia declined overall and Russian entities were suspended from Horizon Europe, the intensity of collaboration in 2024 remains above the expected level for EU-14 and more than 2.5 times the expected level for EU-13. Even though articles stemming from prior projects were published with a temporal lag, the observed intensity of collaboration remains high. Arguably, this can be because Russian scholars remain involved in multilateral projects, albeit without receiving EU funding. This persistence indicates that scientific collaboration between the EU and Russia remained resilient following Russia's full-scale invasion of Ukraine. This aligns with recent studies (Zhang et al., 2024; Plastun et al., 2024).

**Declarations**

**Competing interests** The author has no conflict of interests in the subject matter or materials discussed in this manuscript.

**Acknowledgement**

I am grateful to Tim Engels for insightful comments and suggestions to this article.